\documentstyle[sprocl]{article}

\def\be{\begin{equation}}
\def\ee{\end{equation}}
\def\bea{\begin{eqnarray}}
\def\eea{\end{eqnarray}}

\begin{document}

\title{Theoretical status of the lifetime predictions: \\
       $(\Delta \Gamma/ \Gamma)_{B_s}$, $ \tau_{B^+} / \tau_{B_d} $ 
       and $ \tau_{\Lambda_b} / \tau_{B_d}$}

\author{Alexander Lenz}

\address{Fakult\"at f\"ur Physik, Universit{\"a}t Regensburg, 
         D-93040 Regensburg, Germany
         \\email: alexander.lenz@physik.uni-regensburg.de}


\maketitle\abstracts{We give a review of the theoretical status of the 
         lifetime predictions in the standard model. In case of 
         $(\Delta \Gamma/ \Gamma)_{B_s}$ we are already in a rather advanced 
         stage. We obtain $(\Delta\Gamma/\Gamma)_{B_s}=(9.3^{+3.4}_{-4.6})\%$.
         It seems to be difficult to improve these errors 
         substantially. In addition now some experimental results are 
         available. For $ \tau_{B^+} / \tau_{B_d} $ and
         $ \tau_{\Lambda_b} / \tau_{B_d} $ the theoretical status is much 
         less advanced and the discrepancy between experiment and theory 
         still remains. We conclude with a what-to-do-list for theorists.}

\section{Phenomenology}
  Inclusive decays of B hadrons are expected to be theoretically 
  very clean. Thefore they probe an ideal testing ground for our 
  understanding of QCD in weak decays. 
 Unfortunately there are some discrepancies between experiment and 
 theory, not dramatic, which still have to be resolved. The determination 
 of the semileptonic branching ratio still does not agree perfectly with
 the experimental number. Closely related to that problem is the 
 so-called missing charm puzzle \cite{missing,neub}. 
 If one tunes the input parameter to get a better agreement for 
 $n_c$ one obtains with the same parameters worse values for $B_{sl}$. 
 The present status of the missing charm puzzle can be found in  
 \cite{somecom}. Here clearly more work has to be done.
 The biggest problem in inclusive decays is still the lifetime of the 
 $\Lambda_b$ baryon.
 In the following we discuss the theoretical status of lifetime 
 predictions of B hadrons in the standard model, in particular 
 the decay rate difference $(\Delta \Gamma/ \Gamma)_{B_s}$
 and the lifetime ratios 
 $ \tau_{B^+} / \tau_{B_d} $ and $ \tau_{\Lambda_b} / \tau_{B_d}$
 \footnote{A more detailed discussion can be found in \cite{durham} and
           references therein.}.
\subsection{ The decay rate difference $(\Delta \Gamma/ \Gamma)_{B_s}$}
    Now first experimental numbers for the width difference 
    in the $B_s$ system are available from LEP and CDF \cite{workinggroup}:
    \begin{displaymath}
    \left( \frac{\Delta \Gamma}{\Gamma} \right)_{B_s} = 0.24 ^{+0.15}_{-0.13}
    \; \; \mbox{or} \; \;
    \left( \frac{\Delta \Gamma}{\Gamma} \right)_{B_s} < 0.52 
    \; \; \mbox{at} \; \; 95 \% C.L.
    \end{displaymath}
    If one uses in the experimental analysis the theoretical 
    motivated constraint { $ 1/\Gamma_s = \tau_{B_d}$}, one gets
    a central value of $16 \%$ with an error of $\pm 9 \%$.
    This limits are already very close to the theoretical expectation
    \cite{dgbsnlou}.
    $\Delta \Gamma_{B_s}$ will be measured 
    quite precisely in the near future at TeVatron \cite{rob}.
      Several factors contribute to the big interest in $\Delta \Gamma_{B_s}$:
      a large value of the width difference opens up the possibility for
      novel studies of CP violation without the need for tagging
      \cite{notagging}.
      Moreover, an experimental value of $\Delta \Gamma_{B_s}$ would give
      information about the still unknown mass difference in the $B_s$ system 
      \cite{dgdm} \footnote{The experimental status of $\Delta M_s$ is 
      reviewed in \cite{peter}.}.     
      Another interesting point is 
      that new physics can only lead to a decrease of the width difference
      compared to the standard model value \cite{grossman}. An experimental 
      number which is
      considerably smaller than the theoretical lower bound, would
      thus be a hint for new physics that affects $B_s$-$\bar{B}_s$ mixing
      \footnote{A detailed discussion about new physics effects in the $B_s$ 
      system can be found in \cite{uli1}.}.
      Besides the need for a reliable theoretical prediction of 
      $\Delta \Gamma_{B_s}$ in order to fulfill the above physics program 
      it is of conceptual interest to compare experiment and theory in order
      to test local quark-hadron duality, which is the underlying assumption 
      in calculating heavy quark decay rates. 
      One can show  that duality holds exactly in the limit 
      $\Lambda_{\rm QCD} \ll m_b - 2 m_c \ll m_b$ and $ N_c \to \infty$ 
      \cite{aleksan}. So far no deviation from duality has been 
      conclusively demonstrated experimentally and theoretical 
      models of duality violation in $B$ decays tend to predict 
      rather small effects \cite{uraltsev}.
\subsection{The lifetime ratios: $ \tau_{B^+} / \tau_{B_d} $ and
         $ \tau_{\Lambda_b} / \tau_{B_d} $}
The lifetime ratios of the B-hadrons are quite well known 
experimentally \cite{lifetimeex}.
\begin{displaymath}
     \frac{\tau_{B^+}}{\tau_{B_d}} =  1.074 \pm 0.028 \; , \; \; \;
     \frac{\tau_{B_s}}{\tau_{B_d}} =  0.951 \pm 0.037 \; , \; \; \;
     \frac{\tau_{\Lambda_b}}{\tau_{B_d}} = 0.784 \pm 0.033 \; .
\end{displaymath}
BaBar and Belle expect to pin down the error of the first ratio to 
{$  \pm 1 \% $} \cite{lydia}, BaBar has already a new result:
$ 1.082 \pm 0.028$ \cite{babar}. The lifetimes of the heavier $B_s$ meson
and the $ \Lambda_b$ hadron will be measured at CDF \cite{rob}. 
Theoretically one expects all these ratios to 
be very close to unity, which is clearly not the case.
\section{Theoretical status}
In the framework of the Heavy Quark Expansion (HQE) 
\footnote{For a review see e.g. \cite{hqe}.} 
one can expand the decay rate in inverse powers of the heavy quark mass
\begin{displaymath}
\Gamma  =  \Gamma_0 
+ \left( \frac{\Lambda}{m_b} \right)^2 \Gamma_2
+ \left( \frac{\Lambda}{m_b} \right)^3 \Gamma_3
+ \left( \frac{\Lambda}{m_b} \right)^4 \Gamma_4
+ \cdots \; \; .
\label{hqe}
\end{displaymath}
The leading term is described by the decay of a free
quark and the first non-perturbative corrections arise at the
second order in the expansion.
$\Gamma_0$ and $\Gamma_2$ cancel out in the decay rate difference 
and in the lifetime ratios (except for $\Lambda_b$) and will therefore not 
be discussed in more detail.
In the third order we get the so-called weak
annihilation and pauli interference diagrams. 
Here the spectator quark is included for the first
time. These diagrams give rise to different lifetimes 
for different $B$ hadrons.
Schematically one can write the $\Gamma_i$'s as products of perturbatively
calculable functions (Wilson coefficients) and matrix 
elements, which have to be determined by some non-perturbative methods 
like lattice QCD or sum rules. 
\subsection{ The decay rate difference $(\Delta \Gamma/ \Gamma)_{B_s}$}
The decay rate difference can be written in the following way
\begin{displaymath}
\Delta \Gamma = \frac{\Lambda^3}{m_b^3} \left[
\left( \Gamma_3^{(0)} + \frac{\alpha_s}{4 \pi} \Gamma_3^{(1)} + \dots 
\right) +
\frac{\Lambda}{m_b} \left(\Gamma_4^{(0)} +\dots + \right) + \dots
\right] \; .
\end{displaymath}
$\Gamma_3^{(0)}$ was calculated in \cite{dgbslo} long time ago, the $1/m_b$
corrections were done in $\Gamma_4^{(0)}$ \cite{bbd} and the NLO QCD 
calculation was presented in $\Gamma_3^{(1)}$ \cite{dgbsnlo}.
Both corrections gave a large reduction of the theoretical result.
Meanwhile many  unquenchend ($N_f = 2$) lattice calculations 
of the non-perturbative constants ($f_{B_s}$, $B$ and $B_S$)
which appear in $\Gamma_3$ were done in \cite{dglajap,dglait,dglaes}.
\footnote{The advanced stage of lattice calculations can be read off from a
 comparison of lattice results for $f_{B_d}^2 B_{B_d}$ 
with fits of these parameters from the unitarity triangle \cite{stocchi}.}
The non-perturbative parameters which enter $\Gamma_4$ are still unknown.
In \cite{bbd} the effect of these parameters was estimated in vacuum 
insertion approximation.
\\
The improvement in theory input motivated the update \cite{dgbsnlou}
of the result for $\Delta \Gamma_{B_s}$ presented in \cite{dgbsnlo}. 
We also clarified the origin of seemingly disagreeing recent 
evaluations of $\Delta \Gamma_{B_s}$:
The authors of \cite{dglait,dglaes} were introducing a different
normalization of $\Delta \Gamma/ \Gamma$ in order to get rid of the 
dependence on $f_{B_s}^2$ with the price of getting a strong dependence
on the relatively unknown CKM-parameter $|V_{ts}/V_{td}|$. 
With that method one gets a central value for $(\Delta \Gamma/ \Gamma)_{B_s}$ 
of about $5 \%$.
We \cite{dgbsnlo,dgbsnlou} were normalizing the relative decay 
width difference to the semileptonic 
branching ratio, which should be theoretically very well understood 
\footnote{In principle there could be a similar problem as in the
missing charm puzzle \cite{somecom}, therefore we need precise 
experimental determinations of $B_{sl}$ and $n_c$!}. As lattice calculations
are improving the dependence on $f^2_{B_s}$ should be less important in future.
With the values $f_{B_s} = (230 \pm 30)$ MeV, $B(m_b) = 0.9 \pm 0.1$ and 
$ \bar{B}_S = 1.25 \pm 0.1$ we get as a final number
     \begin{displaymath}
    \left( \frac{\Delta \Gamma}{\Gamma} \right)_{B_s} = 
\left( 9.3 ^{+3.4}_{-4.6} \right) \%
    \end{displaymath}
which coincides with the most recent determination in \cite{dglajap}.
\subsection{The lifetime ratios: $ \tau_{B^+} / \tau_{B_d} $ and
         $ \tau_{\Lambda_b} / \tau_{B_d} $}
The lifetime ratio of two $B$ mesons can be written in the following way 
\footnote{For the ratio $\tau_{\Lambda_b}/\tau_{B_d}$ we have an additional
$1/m_b^2$ term, which is only known in leading order QCD.}
\begin{displaymath}
\frac{\tau_1}{\tau_2} = 1 + \frac{\Lambda^3}{m_b^3} \left[
\left( \Gamma_3^{(0)} + \frac{\alpha_s}{4 \pi} \Gamma_3^{(1)} + \dots 
\right) +
\frac{\Lambda}{m_b} \left(\Gamma_4^{(0)} +\dots + \right) + \dots
\right] \; .
\end{displaymath}
Here the situation is very different. Only
$\Gamma_3^{(0)}$ \cite{lifetime,neub} is known. NLO corrections, 
both in $1/m_b$ and $\alpha_s$ still are missing.
The non-perturbative matrix elements in $\Gamma_3$ 
for the meson lifetime ratio were calculated with
QCD sum rules \cite{china,defazio} and in quenched lattice QCD
\cite{dipierro}.
For the $\Lambda_b$ hadron only preliminary lattice studies are available 
\cite{dipierro2}. In a recent review \cite{flynn} the following 
theoretical numbers were given
\begin{displaymath}
     \frac{\tau_{B^+}}{\tau_{B_d}} =  1.03 \pm 0.04 \; , \; \; \;
     \frac{\tau_{\Lambda_b}}{\tau_{B_d}} = 0.92 \pm 0.02 \; .
\end{displaymath}
This result shows that the leading $1/m_b^3$ corrections 
to $\tau_{\Lambda_b}$ are sizeable. From $\Delta \Gamma_{B_s}$ we have learnt,
that ${\cal O} (\alpha_s)$ and  ${\cal O} (1/m_b)$ corrections to
$\Gamma_3^{(0)}$ are important. So we still have to wait,
till we can claim the discovery of new physics in the lifetime of the
$\Lambda_b$ baryon.
\section{Prospects for improvement} 
For the lifetime ratios $ \tau_{B^+} / \tau_{B_d} $ and
$ \tau_{\Lambda_b} / \tau_{B_d} $ the next steps are clear:
the calculation of the NLO QCD corrections ($\Gamma_3^{(1)}$)
and the $1/m_b$ corrections ($\Gamma_4^{(0)}$) has to be finished.
In the case of $\Delta \Gamma_{B_s}$ all these corrections were quite sizeable
(about $50\%$ of the LO result!).
For the $\Lambda_b $ baryon a lattice determination of the 
non-perturbative matrix elements in $\Gamma_3$ is still missing and for 
mesons one would like to have unquenched results, too.
\\
The next class of improvements (both for $\Delta \Gamma$ and the lifetime 
ratios) could consist of the determination of the  non-perturbative matrix 
elements of the dimension 7 operators, which appear in $\Gamma_4$. 
Morover one could do NLO QCD corrections to the $1/m_b$ corrections 
($\Gamma_4^{(1)}$). As $\Gamma_4^{(0)}$ with vacuum insertion approximation
for the bag parameters is very sizeable in the case of 
$\Delta \Gamma_{B_s}$ \cite{bbd}, 
this effort could be really worth doing it.
Of course, the errors for the decay constants and the bag parameter
will become smaller in future lattice simulations.
To clarify definitly the problem of getting different results for  
$\Delta \Gamma/ \Gamma$ from different normalizations it would be very 
helpful to have precise numbers for $B_{sl}$ from experiment. 
\\
Finally, there are some more possibilities which seem to be quite 
unprobable, to be done in the near future: the
NLO QCD corrections to $\Gamma_2$ for $\Lambda_b$ and the
NNLO QCD corrections to $\Gamma_3$, which would reduce the 
sizeable $\mu$-dependence of $\Gamma_3^{(1)}$.

{\bf Acknowledgments}
I would like to thank the organizers of the workshop for their successful 
work, DFG for financial support and M. Beneke, G. Buchalla, C. Greub and U. 
Nierste for collaboration.

\end{document}